\documentclass{aa}  
\usepackage{graphicx}
\usepackage{txfonts}
\usepackage{hyperref}
\usepackage{amsmath}
\usepackage[separate-uncertainty=true, multi-part-units=single]{siunitx}
\usepackage{comment}

\def\ltsima{$\; \buildrel < \over \sim \;$}
\def\simlt{\lower.5ex\hbox{\ltsima}}
\def\gtsima{$\; \buildrel > \over \sim \;$}
\def\simgt{\lower.5ex\hbox{\gtsima}}
\def\gsimeq
{\hbox{\raise0.5ex\hbox{$>\lower1.06ex\hbox{$\kern-1.07em{\sim}$}$}}}
\def\lsimeq
{\hbox{\raise0.5ex\hbox{$<\lower1.06ex\hbox{$\kern-1.07em{\sim}$}$}}}

\def\erosita{{\it eROSITA}}

\def\efeds{{eFEDS}}
\def\gaia{{\it GAIA}}

\def\apj{ApJ}

\def\mnras{MNRAS}
\def\aap{A\&A}

\def\apjl{ApJ}
\def\apjs{ApJS}

\def\pasj{PASJ}
\def\nat{Nature}

\def\os{O {\sc vii}}

\begin{document}

   \title{On the patchy appearance of the circum-Galactic medium and the influence of foreground absorption}

   \author{G. Ponti\inst{1,2} 
   \and
   J. S. Sanders\inst{2}
   \and
   N. Locatelli\inst{2}
   \and
   X. Zheng\inst{2}
   \and
   Y. Zhang\inst{2}
   \and
   M. Yeung\inst{2}
   \and
   M. Freyberg\inst{2}
   \and
   K. Dennerl\inst{2}
   \and
   J. Comparat\inst{2}
   \and
   A. Merloni\inst{2}
   \and
   E. Di Teodoro\inst{3,4}
   \and
   M. Sasaki\inst{5}
   \and
   T. H. Reiprich\inst{6}
   }
   \institute{INAF-Osservatorio Astronomico di Brera, Via E. Bianchi 46, I-23807 Merate (LC), Italy \\
        \email{gabriele.ponti@inaf.it}
    \and
    Max-Planck-Institut f{\"u}r extraterrestrische Physik, Giessenbachstrasse, D-85748, Garching, Germany 
    \and
    Department of Physics \& Astronomy, Johns Hopkins University, Baltimore, MD 21218, USA
    \and
    Space Telescope Science Institute, 3700 San Martin Drive, Baltimore, MD 21218, USA
    \and
    Dr. Karl Remeis Observatory, Erlangen Centre for Astroparticle Physics, Friedrich-Alexander-Universit\"{a}t Erlangen-N\"{u}rnberg, Sternwartstrasse 7, 96049 Bamberg, Germany
    \and
    Argelander-Institut f\"ur Astronomie (AIfA), Universit\"at Bonn, Auf dem H\"ugel 71, 53121 Bonn, Germany
    }

   \authorrunning{G. Ponti et al.}

   \date{Received \today; accepted future}
 
  \abstract{Recent studies have demonstrated that the emission from the circum-Galactic medium displays a relatively high degree of patchiness on angular scales of $\sim10^\circ$.}
  {Taking advantage of the Spectrum Roentgen Gamma \erosita\ Final Equatorial Depth Survey, we aim to constrain any variation in the X-ray surface brightness on scales going from sub-degree to a hundred square degrees.}
  { } 
  {The surface brightness in the soft X-ray band (0.3--0.45 keV) shows modulations of about $60$~\% on scales of several degrees. The amplitude of such variations decreases for higher energies. The observed patchiness is stable over a period of two years, therefore excluding that it is induced by Solar wind charge exchange. 
  We also observe no correlation between such excess and the density of galaxies in the Local Universe, suggesting no strong contribution from the hot baryons in the filaments of the Cosmic web.
  Instead, the soft X-ray emission is anti-correlated with the column density of absorbing material. Indeed, we can reproduce the spectrum of the bright and dark regions by simply varying the column density of the matter absorbing the emission components located beyond the Local Hot Bubble, while no modulation of the intrinsic emission is required. At high Galactic latitudes, the \erosita\ all sky map shows patchiness of the soft X-ray diffuse emission similar to the one observed in the \efeds\ field, it is therefore likely that the same "absorption-modulation" is present over the entire sky.  }
  {The observed patchiness of soft X-ray diffuse emission within the \efeds\ field is primarily a consequence of absorption. Our spectral decomposition of the soft X-ray background appears accurate, predicting that, apart from the Local Hot Bubble, all other spectral components are modulated by clouds beyond $\sim200$~pc from the Sun. 
  These results highlight the importance of an accurate treatment of the absorption effects, to determine the patchiness of the circum-Galactic medium. }
  

   \maketitle
%
%-------------------------------------------------------------------

\section{Introduction} 

The first all sky maps in the soft X-ray band have revealed the presence of hot diffuse plasma over the entire sky (Snowden et al. 1990; 1994; 1997). 
They also showed the strong modulations on the observed soft X-ray emission induced by intervening absorbing clouds (Snowden et al. 2000). 
Deep observations of some of such obscuring clouds have revealed the presence of the Local Hot Bubble (LHB), a bubble filled with hot ($kT\sim0.1$~keV) plasma with a radius of about $\sim200$~pc, which contains the Sun close to its center (Liu et al. 2017; Zucker et al. 2022). 
Modulations in time of the soft X-ray diffuse emission also revealed the emission induced by charge exchange between the ionised particles of the Solar wind and the neutrals flowing through the Solar system (Kuntz 2019; Snowden et al. 2004). 
Additionally, since the nineties, it was shown that a large fraction of the soft X-ray diffuse emission is produced by the hot plasma around our own Milky Way, the so-called circum-Galactic medium (CGM), as well as a contribution due to the extra-galactic Cosmic X-ray background (CXB). 

The detailed analysis of soft X-ray diffuse emission observed by \erosita\ within the Final Equatorial Depth Survey (eFEDS) allowed Ponti et al. (2022) to decompose the soft X-ray background into its components: emission from Solar wind charge exchange; the local hot bubble; the CGM and the extragalactic CXB. 
In particular, Ponti et al. (2022) characterised the physical properties of the CGM showing that both a thermal component, with temperature consistent with the virial temperature of the Milky Way ($kT\sim0.153-0.178$~keV) and low abundances ($Z\sim0.05-0.1$~$Z_\odot$) as well as a hotter ($kT\sim0.7$~keV) component contribute to the observed soft X-ray diffuse emission (Ponti et al. 2022). 

Although it is clear that the majority of the soft X-ray diffuse emission possesses a Galactic origin, it is still debated what percentage of the total observed flux is produced by the hot baryons in the filaments of the Cosmic web (Vazza et al. 2019). 
For example, Ponti et al. (2022) find that a significant contribution to the soft X-ray background due to hot filaments is required, if the CGM abundance are $Z_{CGM}\gg0.1$~$Z_\odot$. 
Indeed, to reproduce the total emission with a CGM component with the commonly assumed abundances ($Z_{CGM}=0.3$~$Z_\odot$), Ponti et al. (2022) had to add a steep ($\Gamma=4.3-4.8$) power law component. 
Corroborating a significant contribution from the warm-hot intergalactic medium, these authors observed that the surface brightness of such additional component is within the (rather large) range of theoretical expectations from the hot and warm-hot intergalactic medium (Ursino et al. 2011; Roncarelli et al. 2012; Vazza et al. 2019).
However, Ponti et al. (2022) observe that the shape of this additional component surprisingly mimic the shape of the CGM continuum. 
Therefore, these authors suggest that the CGM abundance might indeed be low ($Z_{CGM}\sim0.05-0.10$~$Z_\odot$) and the warm-hot intergalactic medium be contributing to the total soft X-ray background, although at a lower surface brightness than the one required by their additional power law. 
What is clear is that at the intersection of the filaments, the hot plasma in the clusters of galaxies is clearly observed (Sarazin 1986). Cosmological simulations suggest that hot baryons in filaments should provide a contribution to the flux in the soft X-ray band and to its patchiness, however we currently still miss a limpid understanding of the contribution  that we should expect. 
Early \erosita\ (Predehl et al. 2021) observations have already revealed filaments connecting clusters (Reiprich et al. 2022). 
Therefore, deep \erosita\ surveys might have the power to shed light on the emission from hot baryons in filaments (Vazza et al. 2019). 

By analysing the diffuse emission observed by the HaloSat instrument, Kaaret et al. (2020) have reported strong evidence for patchiness of the soft X-ray diffuse emission. 
Such patchiness can be easily understood if either the soft X-ray diffuse emission is produced in a dis-homogenous corona, where the properties of the hot plasma strongly depend on the distribution of heating sources in the underlying disc (i.e., the coronal emission suggested in Ponti et al. 2022). 
Alternatively, the patchiness might be produced by other effects like hot baryons in the filaments or it might be the consequence of the time-variable Solar wind charge exchange emission (SWCX). 
On the contrary, such clumpiness might appear dissonant with the idea of a virialised halo, which is expected to be homogeneous and to cool on a very long time-scale, comparable to the Hubble time. However, second order effects might still produce clumpiness also in an virialised halo. 

During its Calibration and Performance Verification (PV) phase, \erosita\ scanned an extragalactic region of $\sim140$ square degrees to an unvignetted exposure depth of $\sim2.1$~ks ($\sim1.2$~ks vignetted), the so called eFEDS field (Brunner et al. 2021). 
We have studied the spectrum of the diffuse emission, decomposing it in emission from the LHB, the CGM, the Galactic corona and the emission from the extra-galactic CXB (Ponti et al. 2022). 
In this work, we report the detection of patchiness of the soft diffuse emission, on scales of several degrees, within the \efeds\ field. Additionally, we demonstrate that such patchiness is due to the varying column density of absorbing material, therefore excluding that, along these lines of sight, such patchiness is intrinsically due to the CGM. 

\section{Data reduction} 

We analysed the data from the eFEDS field. We followed the same steps and performed the same choices as Ponti et al. (2022), therefore we detail here only the differences compared with that work. 
In particular, as in Ponti et al. (2022), we used only the "on chip" filter cameras, however we summed the spectra of the different instruments and fitted the combined spectrum. 
Data are processed and analysed with the version eSASSuser\_020 of the \erosita\ Science Analysis Software System (eSASS)\footnote{Please, visit the page: https:\/\/erosita.mpe.mpg.de\/edr\/DataAnalysis for more information about eSASS.}. 
Images are displayed with a pixel size of 16 arcseconds. 
As in Ponti et al. (2022), we refer to the PV phase observations as e0 and the first \erosita\ All Sky Survey (eRASS1), eRASS2, and eRASS3 as e1, e2, and e3, respectively, while e12 corresponds to the sum of e1 and e2. 

By investigating the \erosita\ maps in the 2.3--4.5~keV band, which are dominated by particle background, we observe one episode\footnote{An enhancement in the flux of particle background can be observed in the map as a bright stripe, as a consequence of the scanning pattern of the telescope. Indeed, during the all sky survey, \erosita\ continuously performs great circles passing through the Ecliptic poles and advancing by about 1 degree per day (Predehl et al. 2021). } during which the particle background is enhanced by $\sim80$~\% during e3 in the 2.3--4.5 keV band and $\sim30$~\% in the 0.3--1.4~keV band. 
Such enhancements are often associated with mass ejections of the corona of the Sun. These coronal mass ejections then induce a temporary enhancement of the particle background (more prominent at higher energies) observed by the \erosita\ detectors. 
We take into account of this excess emission by properly re-normalising the particle background level in all our spectral fits. 
Additionally, we have chosen to perform our analysis within the 0.3--1.4~keV energy band, so that the particle background provides a minimal contribution to the total emission (see Sect. 4 of Ponti et al. 2022). 

\section{Patchiness of the soft X-ray diffuse emission}

\begin{figure*}[t]
\includegraphics[width=1.0\textwidth]{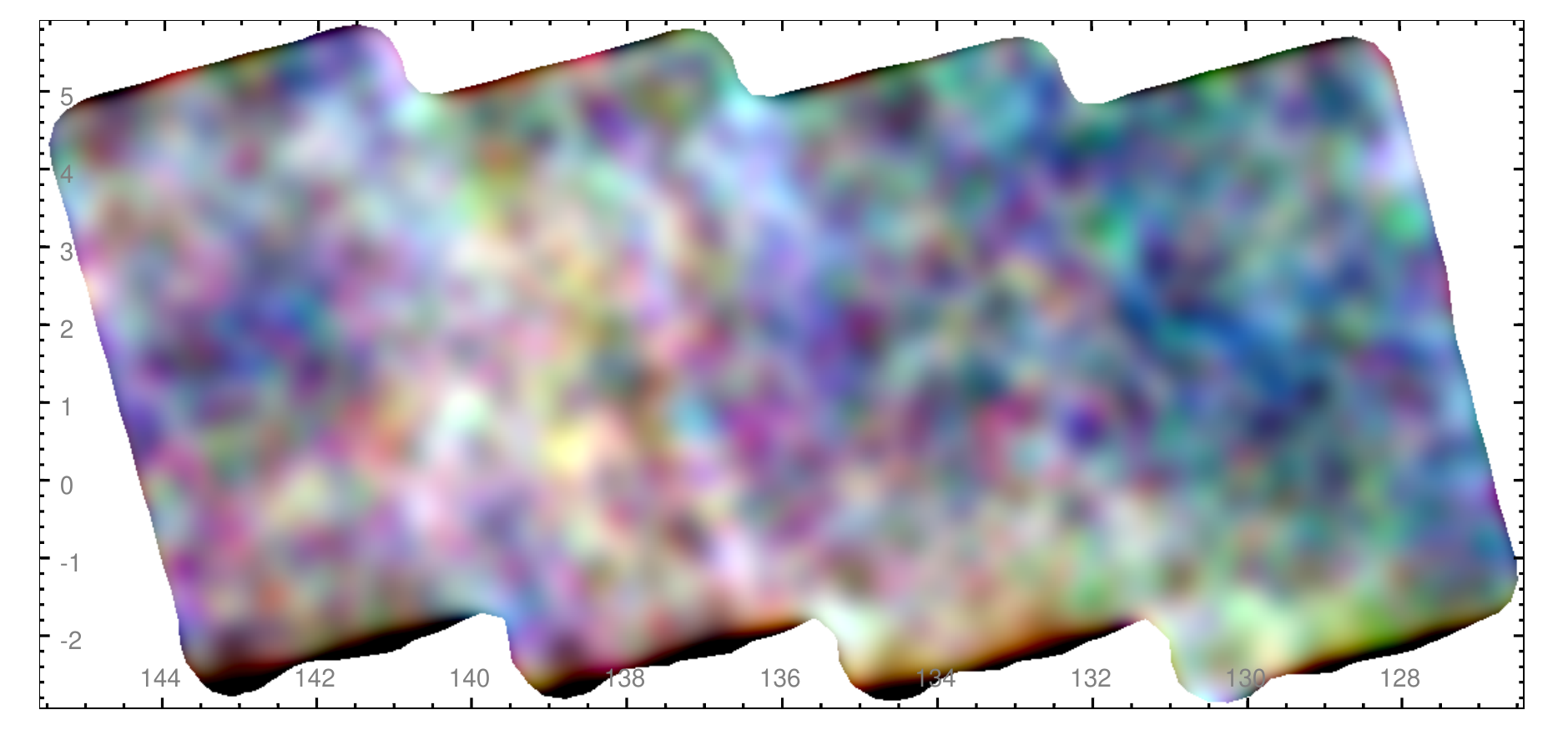}
\caption{Total X-ray emission observed by \erosita\ in the \efeds\ region, during e0. The contribution due to point sources has been removed through a wavelet filtering. With red, green and blue colors are shown the 0.3-0.45~keV; 0.45-0.7~keV and 0.7-2.3~keV bands, respectively. Variations of the surface brightness of the diffuse emission are clearly observed from a sub-degree to several degrees scale. The brightest patch appears slightly redder, consistent with the inference that the flux in the soft band is less absorbed. Equatorial coordinates are used. }
\label{RGBPV}
\end{figure*}
\begin{figure}[t]
\includegraphics[width=0.5\textwidth]{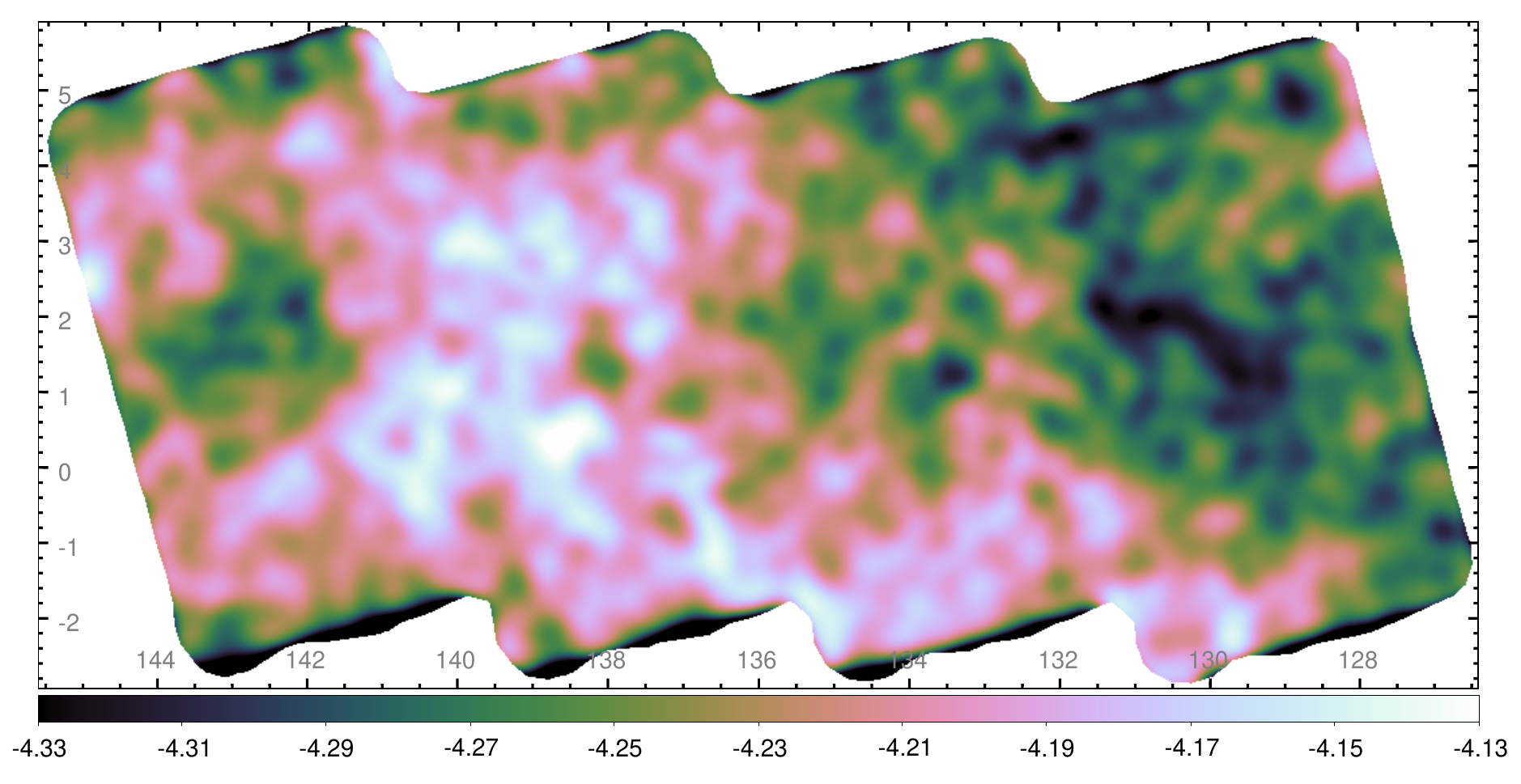}
\includegraphics[width=0.5\textwidth]{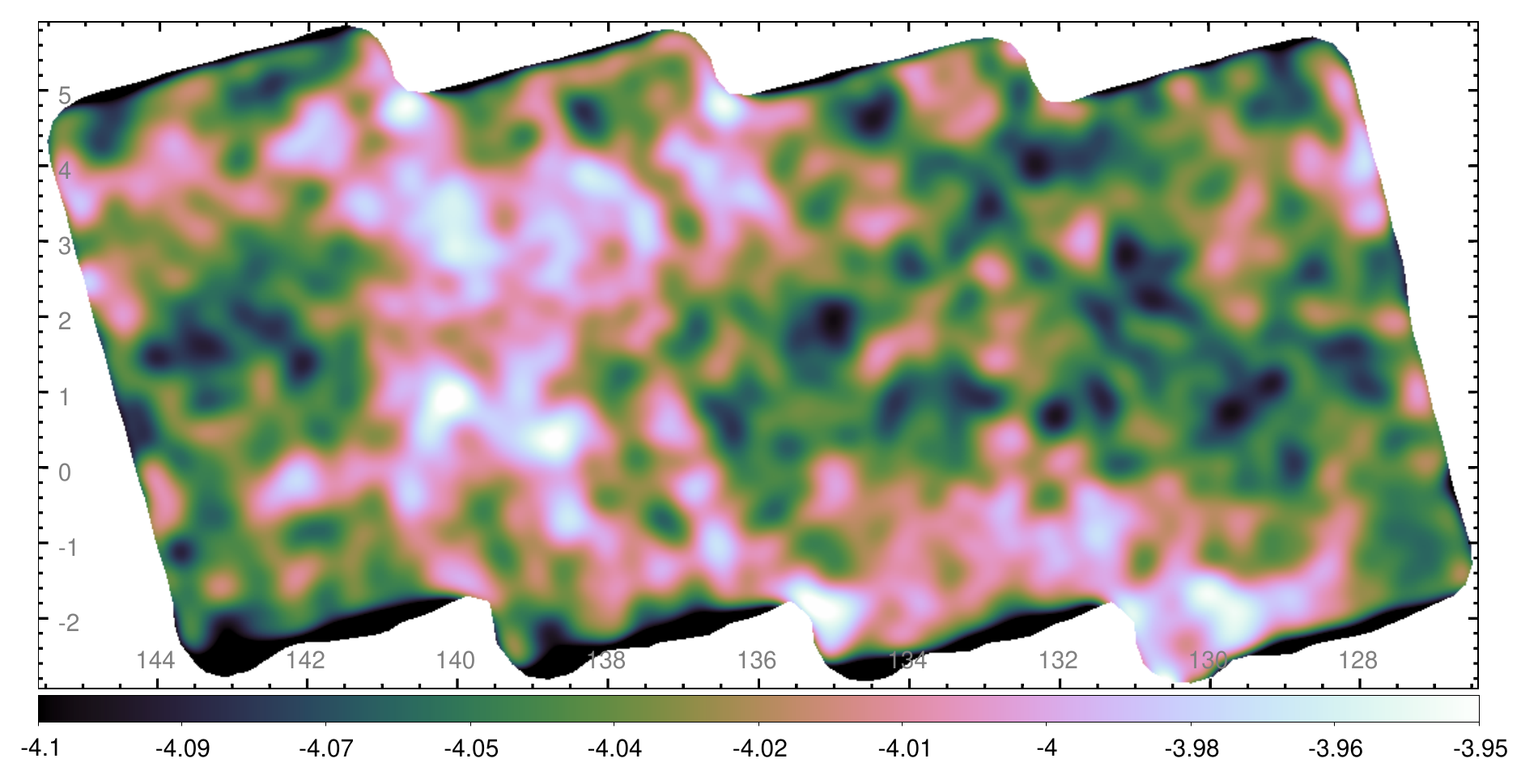}
\includegraphics[width=0.5\textwidth]{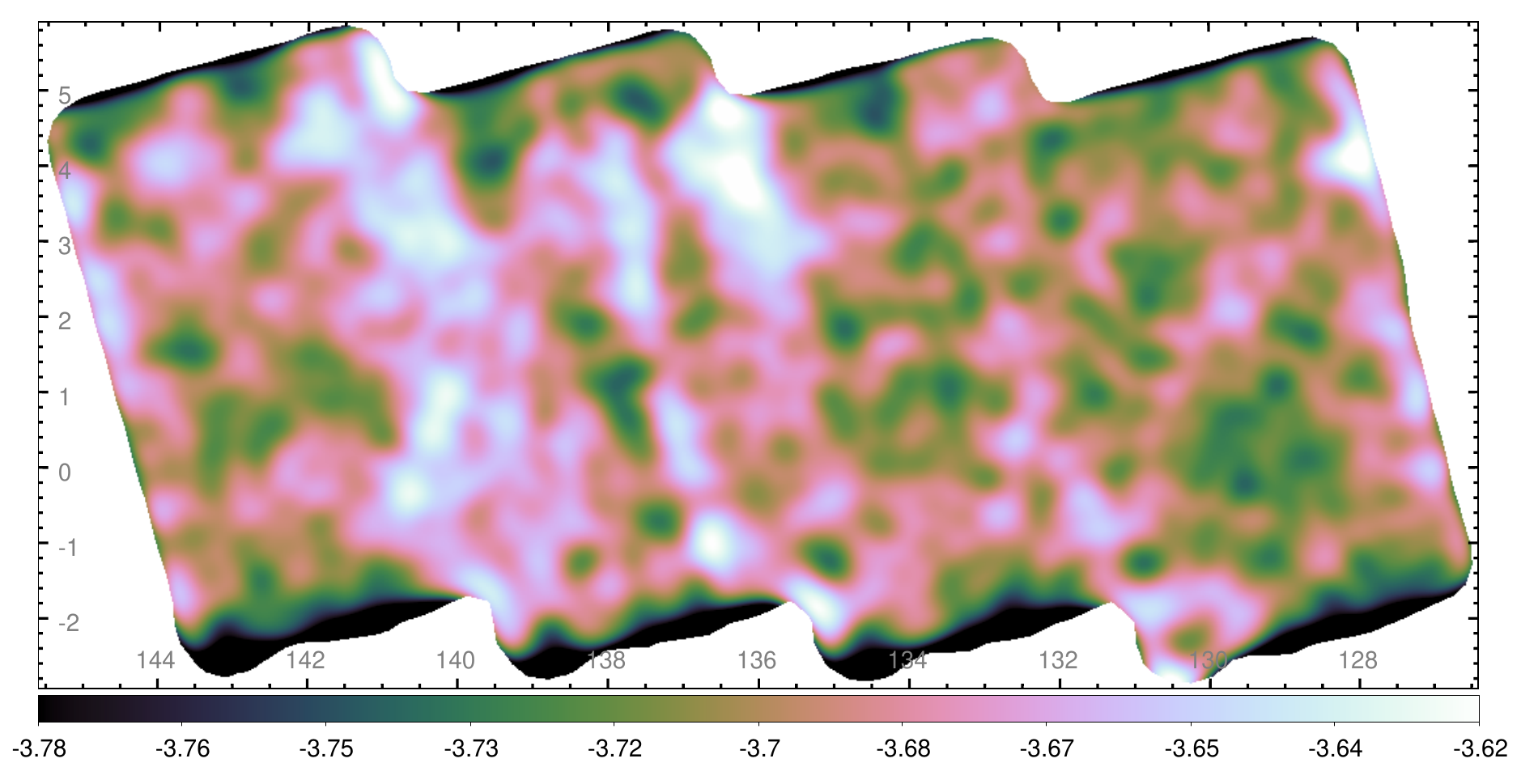}
\caption{From top to bottom the total X-ray emission observed by \erosita\ in the \efeds\ region during e0 is shown in the 0.3-0.45~keV; 0.45-0.7~keV and 0.7-2.3~keV bands, respectively. The contribution due to point sources has been removed through a wavelet filtering. Variations of the surface brightness, as large as a factor of $\sim1.6$, are observed in the 0.3--0.45 keV band, on scales of several degrees. The 0.7-2.3~keV band show few stripes induced by enhanced particle background, as discussed in Sect. 2. Equatorial coordinates are used.}
\label{singlebands}
\end{figure}

The RGB image in Fig. \ref{RGBPV} shows the intensity of the soft X-ray emission as observed by \erosita\ in the eFEDS field during e0, after applying the wavelet filtering to remove point sources and clusters (Vikhlinin et al. 1997)\footnote{In particular, scales of 1 to 512 pixels (up to 8.5 arcmin) were filtered out, using a detection threshold of 3 and a filtering threshold of 2, with 10 iterations (see https://github.com/avikhlinin/wvdecomp). }.
To better appreciate the amplitude of the modulations in the various bands, Figure \ref{singlebands} shows the 0.3--0.45, 0.45--0.7 and 0.7--2.3~keV band maps from top to bottom, respectively. 
The top panel of Fig. \ref{singlebands} shows an enhancement of the soft X-ray background count rate of $cr_{0.3-0.45}\sim5\times10^{-5}$ photons s$^{-1}$ pixel$^{-2}$ around ($\lambda$,$\beta$) $\sim$ (139$^\circ$,1$^\circ$) and a depression of $cr_{0.3-0.45}\sim8\times10^{-5}$ photons s$^{-1}$ pixel$^{-2}$ around ($\lambda$,$\beta$) $\sim$ (130$^\circ$,1$^\circ$). 
The amplitude of such excess drops with energy going from maximum to minimum ratio of $\sim1.6$ in the 0.3--0.45~keV band to $\sim1.2$ in the 0.7--2.3~keV band (Fig. \ref{singlebands}).

In particular, Fig. \ref{singlebands} shows a clear modulation of the brightness on scales going from few square degrees to several square degrees and possibly extending beyond the \efeds\ region. 
Indeed, the brightest patch around ($\lambda$,$\beta$) $\sim$ (139,2$^\circ$) has a projected extension of $\sim5^\circ$ or longer. 

\subsection*{Can the observed patchiness be induced by Solar wind charge exchange (SWCX)?}
\label{SWCX}

As a consequence of its time-variable nature, SWCX could induce patchiness in the observed diffuse emission, if: i) it provides a significant contribution to the total emission, and; ii) the variations in the intensity of the SWCX component happened on a time-scale shorter than the $\sim6$~days that \erosita\ required to scan the \efeds\ region during e0. 

\begin{figure}[t]
\includegraphics[width=0.5\textwidth]{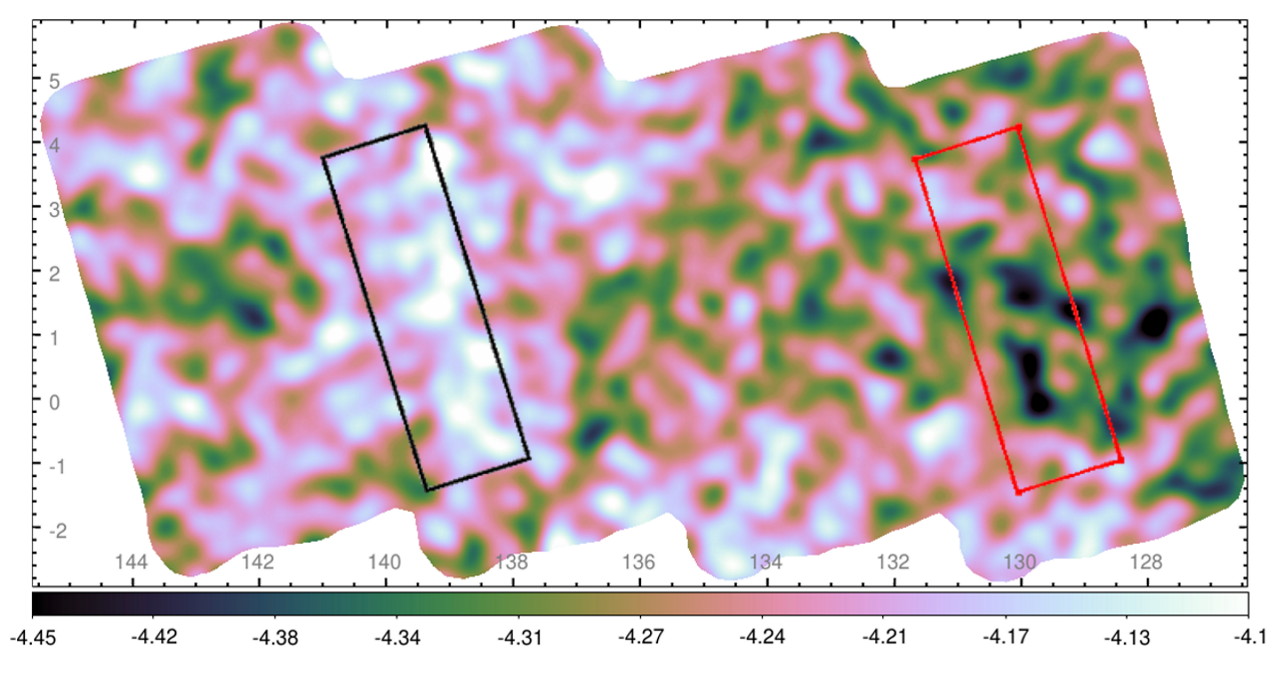}
\includegraphics[width=0.5\textwidth]{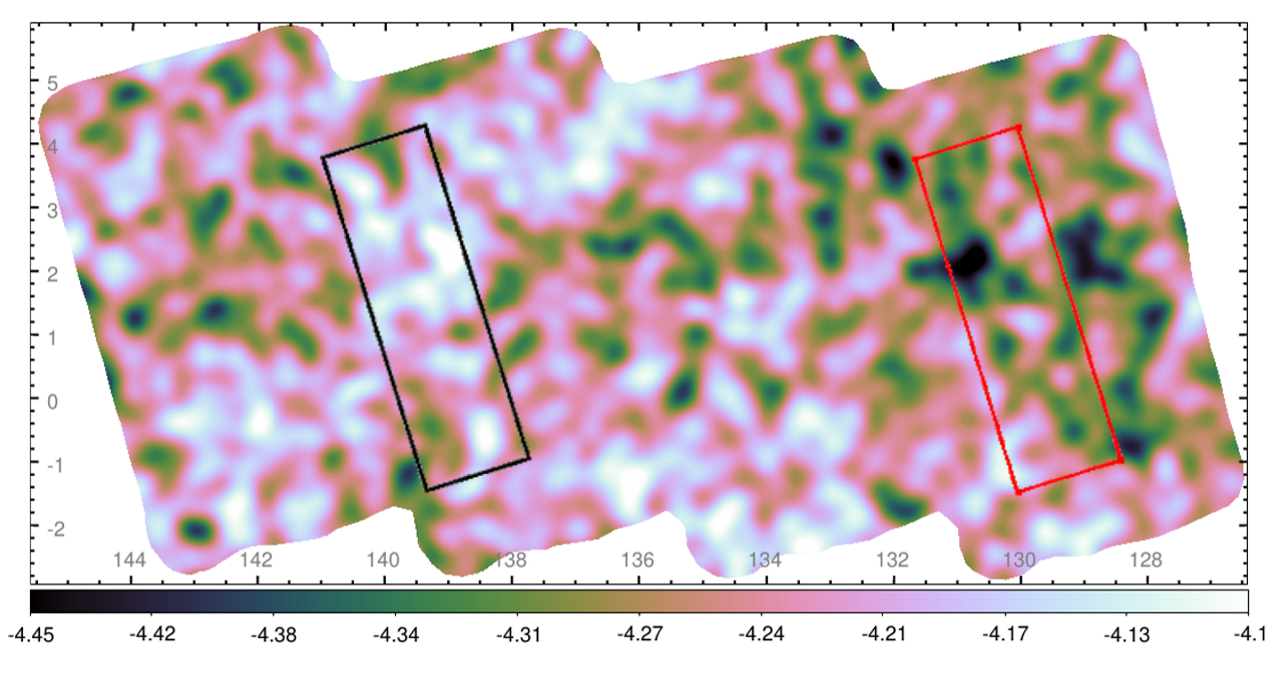}
\caption{Top and bottom panels show the same image as top panel of Fig. \ref{singlebands}, but for e1 and e2, respectively. Despite the worse statistics, the same enhancement of the soft X-ray emission is observed at the same location as during e0 (Fig. \ref{singlebands}). The black and red rectangle show the extraction regions for the bright and dim spectra, respectively. Please, note the slightly different color scale compared with Fig. \ref{singlebands}. Equatorial coordinates are used.}
\label{softXrayeDiff}
\end{figure}

To verify whether this is indeed the case, we investigated the variability in time of the morphology of the observed patchiness. 
Indeed, due to the stochastic nature of the fluctuations of the SWCX emission, if SWCX would be the dominant factor inducing the excess, then its morphology is expected to be different between e0, e1 and e2, which are accumulated about 6 months apart from each other. 
Instead, Fig. \ref{softXrayeDiff} shows that, despite the lower signal to noise due to the shorter exposures in e1 and e2, a consistent excess, with similar morphology as measured during e0, is also observed during e1 and e2. 
Therefore, this excludes that the observed patchiness in the soft X-ray diffuse emission is induced by a time-variable component, such as SWCX. 

\section{Spectrum of the excess emission}
\label{SectDiff}

\begin{figure}[t]
\includegraphics[width=0.5\textwidth]{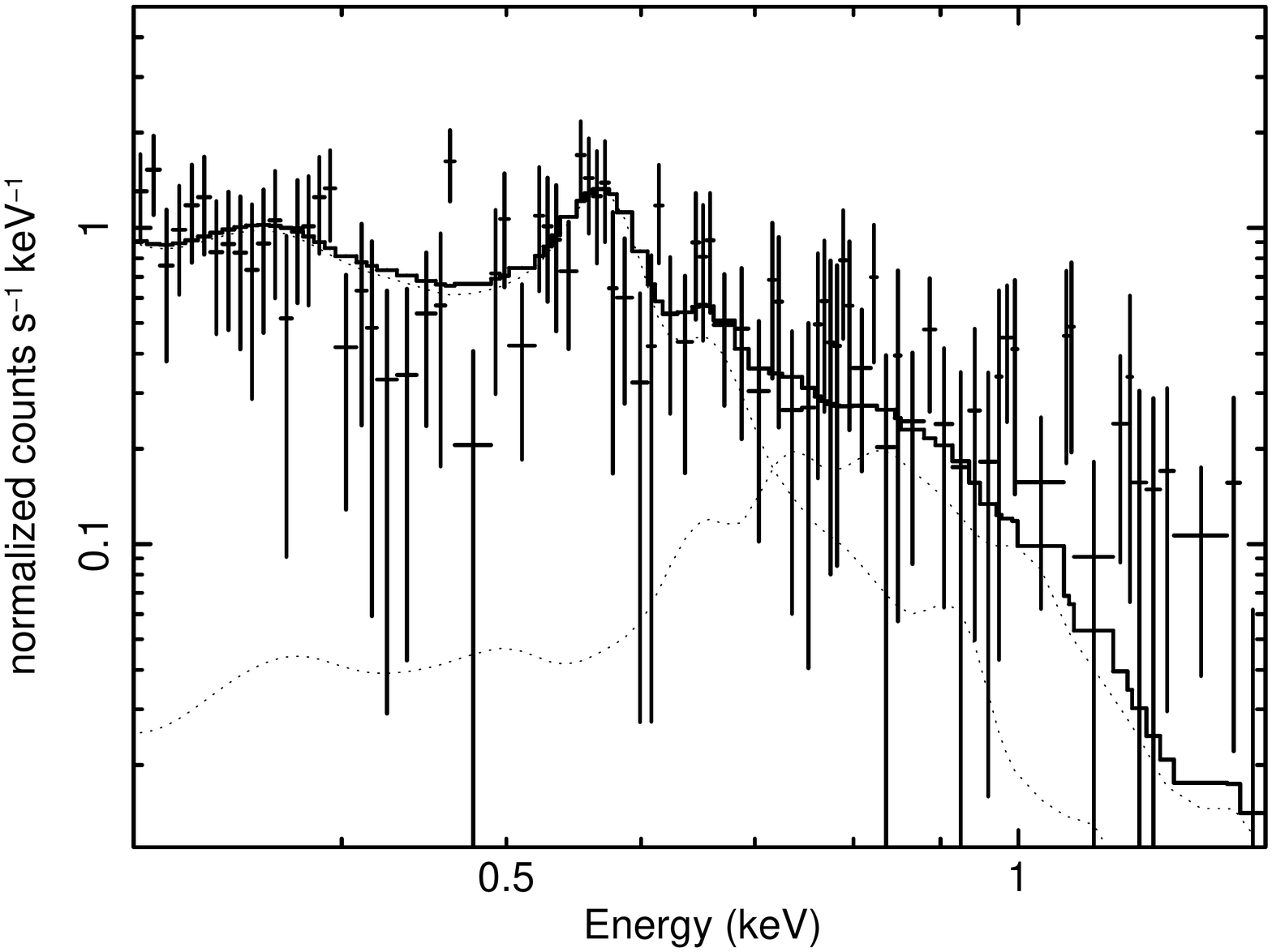}
\caption{Spectrum of the diffuse emission extracted from the black rectangle in Fig. \ref{softXrayeDiff} (called "bright") minus the emission from the red rectangle ("dim"). The black line shows the best fit model, while the dotted lines show the two best fit hot plasma components. At energies higher than $\sim1$~keV the difference spectrum shows a rapidly decreasing spectrum. The plot, in logarithmic scale shows only the the fraction of energy bins with positive values, while the negative values are not displayed. Consequently, at high energies the data displayed are biassed high, while the model properly accounts for both positive and negative energy bins. }
\label{Diff}
\end{figure}

To better investigate the nature of the excess emission, we extracted spectra of the bright and faint emission as shown by the black and red rectangles in Fig. \ref{softXrayeDiff}. Figure \ref{Diff} shows the spectrum of the bright region minus the one of the faint emission (i.e., the total emission from the black minus the one from the red rectangle in Fig. \ref{softXrayeDiff}). Therefore, the difference spectrum shown in Fig. \ref{Diff} reports the spectrum of only the excess component present in the bright region, while all other spatially homogeneous components are subtracted off. 

We start by fitting the difference spectrum with a simple power law model, absorbed by a negligible column density ($Log(N_H/1cm^{-2})=19.5$) of neutral material ({\sc disnht*powerlaw} in {\texttt Xspec})\footnote{The {\sc disnht} model has been developed in order to reproduce interstellar medium absorption in X-ray spectra, extracted over spatial regions encompassing a set of different column densities (Locatelli et al. 2022), such as in the present case. We chose to fix the column density of absorption to $Log(N_H/1cm^{-2})=19.5$, which is the smallest column density allowed by the {\sc disnht} table model.}. 
Despite this model provides an acceptable description of the data ($\chi^2=179.1$ for 162 dof), the best fit slope of the power law results to be rather steep, with a photon index of $\Gamma\sim3\pm0.2$. 
This might indicate that the emission possesses a thermal origin. 
The thermal origin of the underlying emission is further corroborated by the detection of an excess of emission at $E\sim0.56$~keV, which is likely associated with \os\ emission. 
To test this hypothesis, we add to the model a narrow Gaussian line with energy at $E=0.57$~keV ({\sc disnht*(powerlaw+Gauss)} in Xspec). 
We observe a slight improvement of the fit ($\Delta\chi^2=6.8$ for the addition of a new free parameter, corresponding to an F-test probability of $\sim99$~\%). 

Considering the likely thermal nature of the excess emission, we substitute the power law continuum with an optically thin hot plasma component (fitted with an {\sc apec} component in Xspec, therefore resulting in: {\sc disnht*apec}\footnote{Also for this model we assume the column density of absorbing material to be negligible. }). 
This model describes the bulk of the emission with a thermal emission with $kT=0.21\pm0.02$~keV and abundances $A=0.05\pm0.02$ times Solar, however leaving clear residuals above $\sim1$~keV. 
Additionally, this model does not adequately describe the observed spectrum ($\chi^2=185.1$ for 161 dof). 

The observed complexity might be the consequence of emission due to more than one temperature plasma (e.g., Ponti et al. 2022).  
Therefore, we add a second optically thin thermal component to the model ({\sc disnht*(apec+apec)}). 
Considering the low statistics of this difference spectrum, we assumed the metal abundances to be the same for the two components. 
This model provides the best description of the data ($\Delta\chi^2=166.2$ for 159 dof). 
The best fit temperature of the two plasma are $kT_1=0.14\pm0.03$~keV and $kT_2=0.7^{+0.4}_{-0.2}$~keV, with a best fit metal abundance of $A=0.05\pm0.03$~Solar (Fig. \ref{spec}). 
Incidentally, we note that the temperatures and metal abundances of these plasma components are consistent with the ones of the CGM and coronal components observed in the total \efeds\ spectrum (Ponti et al. 2022). 

\subsection{Can the patchiness be induced by hot baryons in the filaments of the Cosmic web?}

We note that, such as observed here, hot baryons along the filaments of the Cosmic web are expected to produce optically thin thermal radiation with temperatures in the range of $\sim0.1$ to $\sim0.7$~keV and with metal abundances of the order of $Z\sim0.05-0.1$~$Z_\odot$ or lower (Vazza et al. 2019). 
Additionally, in this scenario, the emission measure of the soft X-ray diffuse emission should be modulated by the intrinsic structure of the Cosmic web. 
In particular, the imprint of the nearest structure of the Cosmic web might produce patchiness on several degrees scales. 
Therefore, it is expected that the hot baryons in the filaments would induce a patchiness of the soft X-ray diffuse emission, as observed in the \efeds\ field. 

In order to determine whether this is indeed the case, we constrained the distance from which such thermal emission is produced by fitting the redshift of the lower temperature hot plasma component. 
This is possible thanks to the observation of the soft X-ray emission lines (i.e., \os) in the difference spectrum. 
We observe that the redshift is $z=0\pm0.03$ at 90~\% confidence. 
Therefore, if this excess emission is associated with hot baryons in the Universe, then they must be confined within the Local Volume. 

%\subsubsection*{Correlation with nearby galaxies}

\begin{figure}[t]
\includegraphics[width=0.5\textwidth]{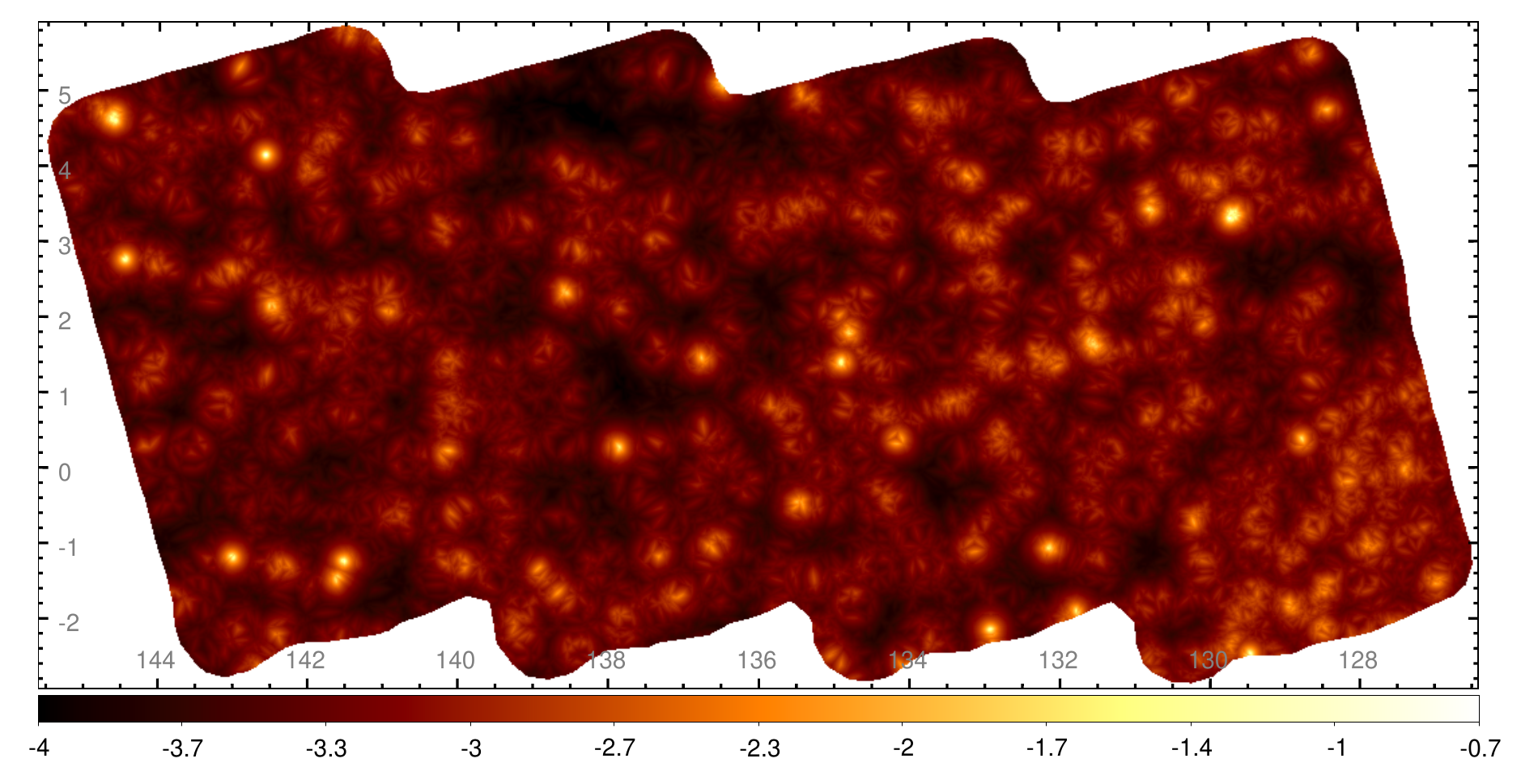}
\caption{Spatial distribution of the logarithm of the number density of galaxies per pixel. Equatorial coordinates are used.}
\label{galaxies}
\end{figure}

To test whether hot baryons can be the origin of the excess emission and the observed patchiness, we need to choose a tracer of the Cosmic web. 
By assuming that galaxies more massive than the Milky Way ($M_\star>6\times10^{10}$~M$_\odot$) are good tracers of the local Cosmic web, we estimate the spatial distribution of the galaxies which are at distances lower than $z<0.03$ and with a stellar mass $M_\star>6\times10^{10}$~M$_{\odot}$. 
To achieve this, we use both the GAlaxy and Mass Assembly (GAMA; Liske et al. 2015; Driver et al. 2022) catalogue of galaxies as well as the Heraklion Extragalactic Catalogue of galaxies (HECATE; Kovlakas et al. 2021). 
The advantage of the latter is that HECATE covers the entire area of the eFEDS field, while GAMA covers only $\sim60$ square degrees (see fig. 1 of Comparat et al. 2022). 
On the other hand, the disadvantage is that HECATE is not a complete sample of galaxies. 
Indeed, we verified this by comparing HECATE and GAMA on the common area. 
To be confident that the results presented here are independent from the catalogue of galaxies employed, we performed the analysis with each sample, obtaining consistent results. 
To further verify this, we used a yet different catalogue. 
Indeed, we selected nearby galaxies from the Legacy photometric survey (DR9; Dey et al. 2019) on the basis of their photometric redshift and color plus magnitude\footnote{We apply the selection: 
$z_p < 0.05 \&\& 22.5 - 2.5 \times log10(F_R) < 20  \&\& 22.5 - 2.5 \times log10(F_R) > 16$, where $z_p$ is the photometric redshift (we allow for a small uncertainty on this value, therefore allowing also values between 0.03 and 0.05) and $F_R$ is the flux in the R band. Additionally, we select: $22.5 - 2.5 \times log10(F_Z) - (22.5 - 2.5 \times log10(F_{W1})) > 22.5 - 2.5 \times log10(F_G) - (22.5 - 2.5 \times log10(F_R)) - 1.5$, where $F_Z$, $F_{W1}$ and $F_G$ are the fluxes in the z, W1 and g bands, respectively, are in order to efficiently remove stars. }. 
In this way we can obtain a rather complete view on the galaxies within the \efeds\ region. The remaining contribution due to nearby stars has been removed by dismissing all sources with {\sc gaia} parallaxes measured at better than 3 sigma. 
Figure \ref{galaxies} shows the distribution of the selected galaxies within the \efeds\ region. We estimate the density of galaxies in each point by applying the nearest neighbour interpolation algorithm\footnote{The algorithm finds the radius around each point which contains N galaxies, then it computes the density given N and the area of that circle. For small numbers of galaxies, the algorithm produces some of the high angular frequency geometric artefacts observed in Fig. \ref{galaxies} }. 

The results obtained with this selection are consistent with the ones obtained by selecting galaxies from the GAMA and HECATE ones. Hereinafter, we show the results of the comparison with the galaxies from the Legacy photometric redshift sample, only. 

\begin{figure*}[t]
\begin{center}
\includegraphics[width=0.8\textwidth]{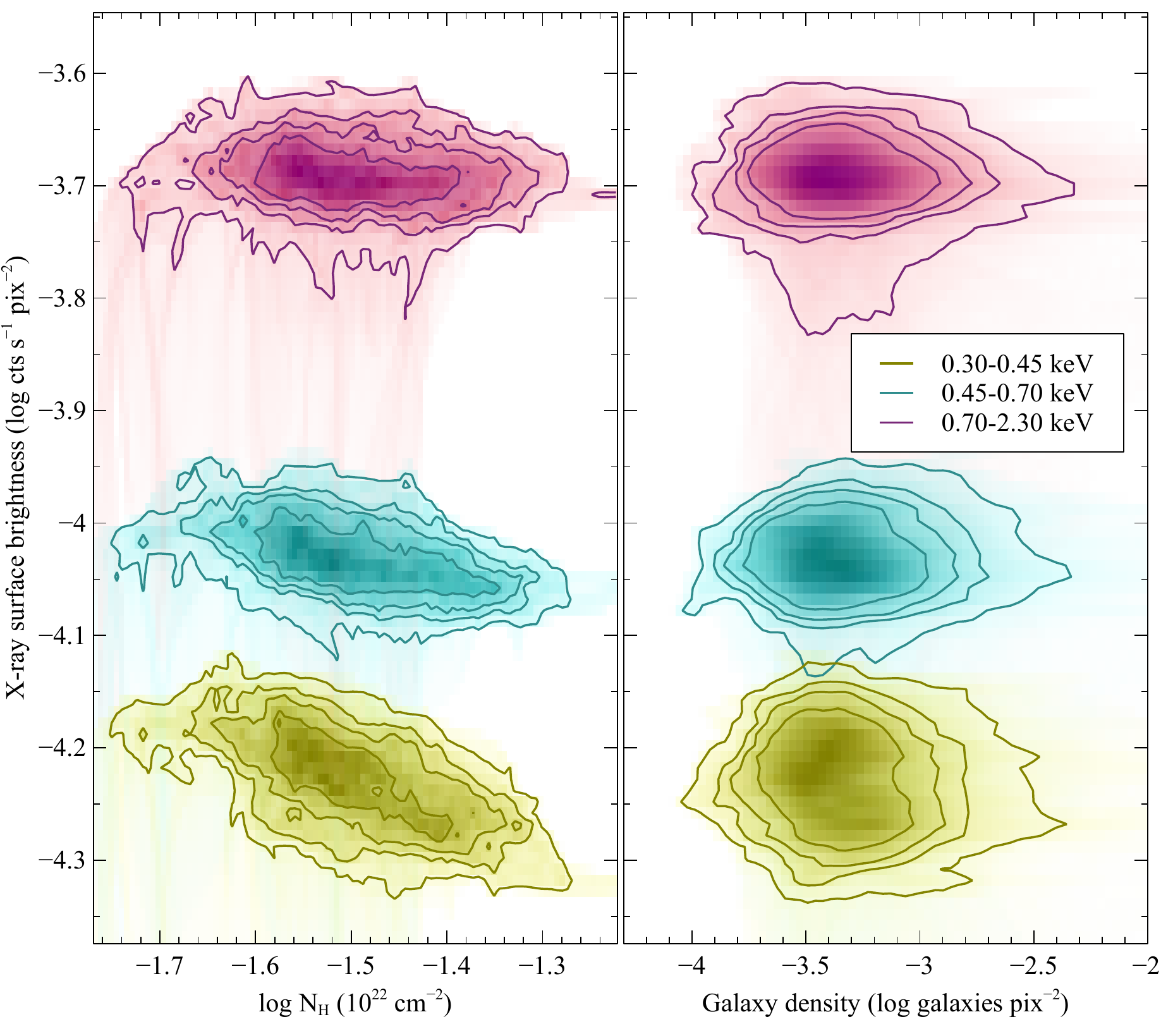}
\caption{{\it (Left panel)} Correlation between the X-ray surface brightness in three energy bands (0.3--0.45, 0.45--0.7, 0.7--2.3~keV in yellow, cyan and red, respectively) during e0 vs. the column density of neutral material. {\it (Right panel)} Correlation between the X-ray surface brightness in three energy bands vs. the density of galaxies which are a proxy for the distribution of hot baryons in Cosmic filaments. }
\label{nhcorr}
\end{center}
\end{figure*}

We then correlate the spatial density of galaxies with the count rate map in the 0.3--0.45~keV energy range, to verify whether the observed patchiness is produced from regions with higher densities of galaxies, therefore from Cosmic filaments (right panel of Fig. \ref{nhcorr}). 
We observe that galaxies are not correlated with the intensity of the diffuse emission observed by \erosita. 
Therefore, we conclude that the patchiness observed in the \efeds\ region is not due to hot baryons in filaments. 

\subsection{Can absorption induce the observed patchiness? }
\label{SectAbs}

\begin{figure}[t]
\includegraphics[width=0.5\textwidth]{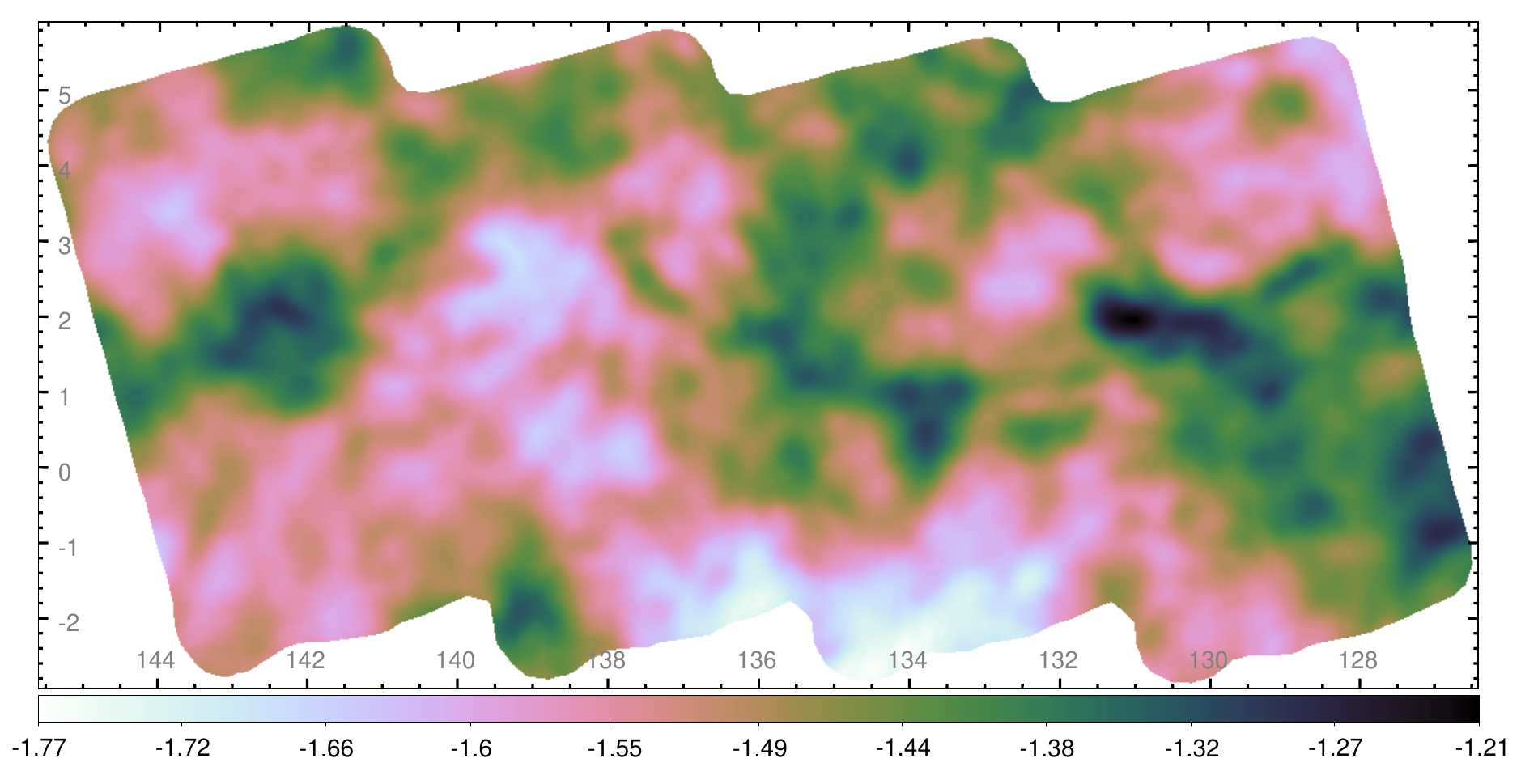}
\caption{Spatial distribution of the logarithm of the column density of neutral absorbing material within the \efeds\ region in units of log(N$_H$/10$^{22}$ cm$^{-2}$). White and pink colours display regions with low column density, while darker colours indicate lines of sight characterised by larger absorption. Equatorial coordinates are used.}
\label{nh}
\end{figure}
Another possible origin for the patchiness of the soft X-ray diffuse emission is the variation of the column density of the intervening matter. 
In this case, we would expect that the difference spectrum would be produced by the same emission components as observed in the total spectrum (Ponti et al. 2022), which would be simply modulated by the variable absorption. 
Indeed, the best fit temperatures and metal abundance of the thermal components fitting the difference spectrum (see Sect. \ref{SectDiff}) are remarkably similar to the ones of the CGM and coronal components observed in the total \efeds\ spectrum (Ponti et al. 2022). 

To further test whether the observed patchiness is induced by absorption, we show in Fig. \ref{nh} the spatial distribution of the column density of neutral absorbing matter, as measured by the NH4PI survey, within the \efeds\ region (please note the inverted colour scale with larger column densities appearing as darker colours). 
Variations as large as a factor of $\sim3$, going from $N_H\sim2\times10^{20}$~cm$^{-2}$ to $N_H\sim7\times10^{20}$~cm$^{-2}$ are observed within the \efeds\ region. 
Indeed, such large amplitude column density variations might have an effect on the observed flux in the $0.3-0.45$~keV energy band. 

This inference is corroborated by the negative correlation between the logarithm of the column density and the surface brightness in the 0.3--0.45~keV band (yellow contours in the left panel of Fig. \ref{nhcorr}). 
This is further strengthened by the energy dependence of such correlation.
Indeed, we observe that the anti-correlation between surface brightness and column density of neutral absorption becomes weaker with increasing energies (see blue and red contours in the left panel of Fig. \ref{nhcorr}). 
This is expected, considered that absorption of X-ray photons by neutral material is exponentially decreasing with energy.  
In particular, Fig. \ref{nhcorr} shows that no significant correlation is observed between the surface brightness in the higher energy band (0.7--2.3~keV) and the column density of absorption. 
This result strongly suggests that the observed patchiness in the \efeds\ region is primarily induced by the different column density of neutral absorption. 

\begin{figure}[t]
\begin{center}
\includegraphics[width=0.5\textwidth]{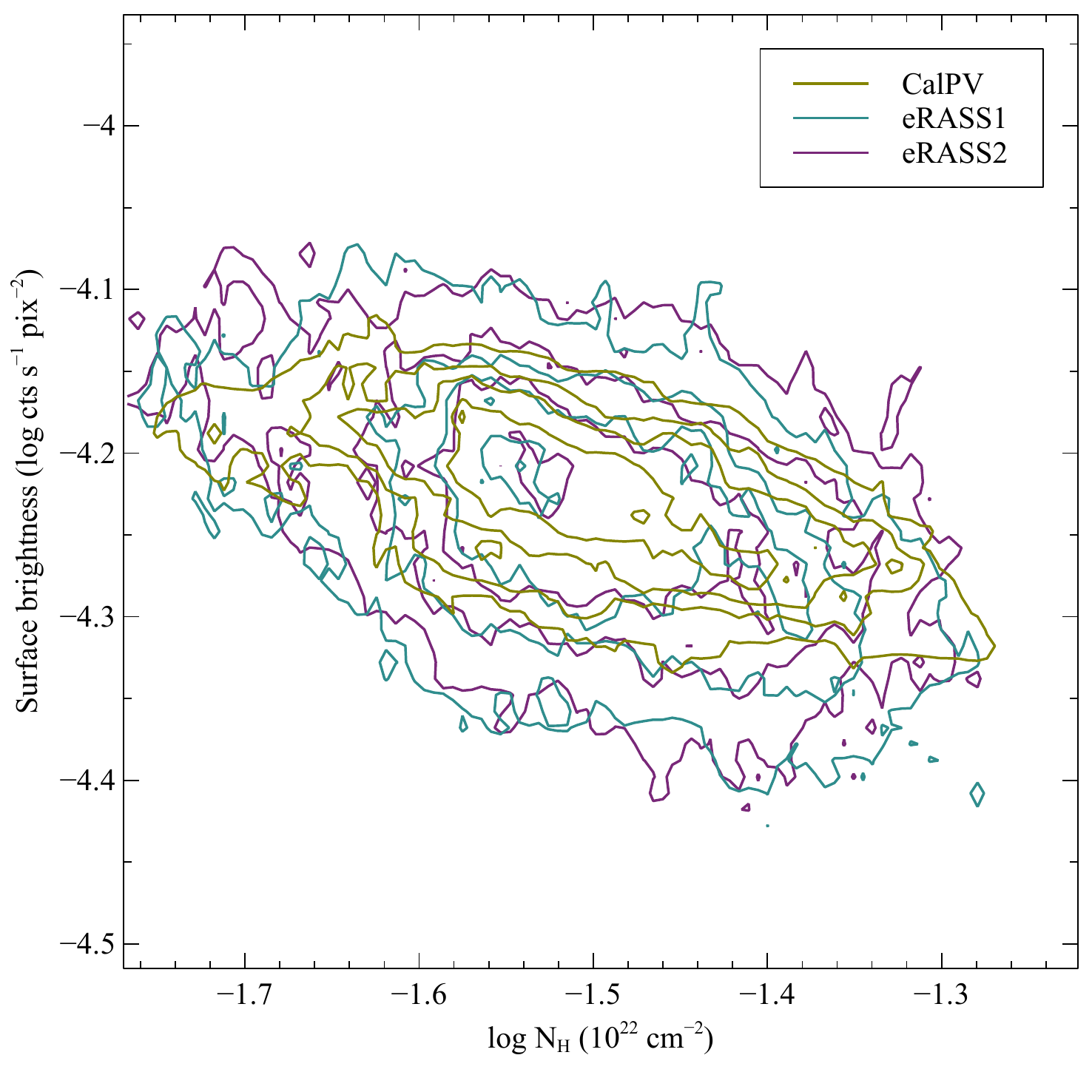}
\caption{Correlation between the X-ray surface brightness in the 0.3--0.45~keV energy band during e0, e1, and e2, in yellow, cyan and magenta, respectively, vs. the column density of neutral material. }
\label{e12}
\end{center}
\end{figure}

Figure \ref{e12} shows that the same correlation holds also during e1 and e2, further strengthening the presence of such correlation and indicating that the effects of SWCX are negligible (Sect. \ref{SWCX}). 

\section{Modulation by absorption: No need for intrinsic patchiness of the CGM}
\label{abs}

To provide further evidence that the observed patchiness is induced primarily by absorption, we fit the bright and dim spectra with the same model used to fit the total \efeds\ spectrum as in Ponti et al. (2022)\footnote{We fitted the model shift-LHB-CGM-Coro2-CXB as defined in that work.}, simply re-normalising the total emission by the ratio of the extraction regions\footnote{The spectrum analysed in Ponti et al. (2022) was extracted from a region of 107.5 square degrees, while the bright and dim regions have been accumulated from regions of 9.35 square degrees, each (see Fig. \ref{softXrayeDiff}). }.
The model is composed by the un-absorbed LHB, the CGM component, the corona and the CXB, absorbed by a column density of neutral material ({\sc apec+disnht*(bkn2pow+apec+rnei} in Xspec). 
We keep all parameters of the model fixed to the best fit values observed in the total \efeds\ spectrum (see model v-shift-LHB-CGM-Coro2-CXB of Table 2 of Ponti et al. 2022), leaving only the column density of neutral material to be free.
\begin{figure}[t]
\vspace{-0.8cm}
\includegraphics[width=0.5\textwidth]{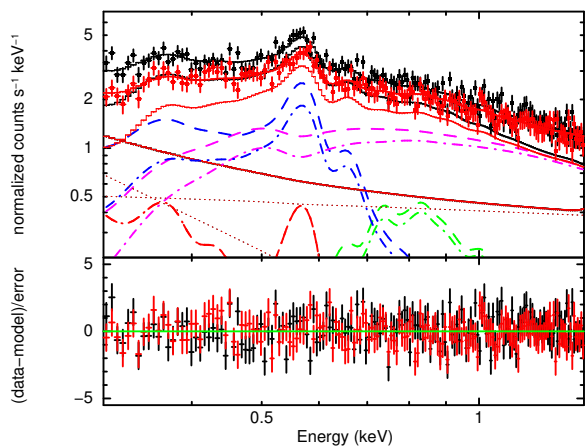}
\caption{The black and red data points show the spectra of the bright and dim emission, respectively, during e12. The red, blue, green, and magenta lines indicate the emission from the LHB, the CGM, the corona and the CXB (dashed and dash-dotted lines for the bright and dim spectra), respectively.}
\label{spec}
\end{figure}

The black and red data in Fig. \ref{spec} show the spectrum of the bright and dim emission as extracted from the regions shown in Fig. \ref{softXrayeDiff}. 
The dashed and dash-dotted lines in Fig. \ref{spec} show the model components for the bright and dim spectra, respectively. 
In red, blue, green, and magenta the emission from the LHB, the CGM, the corona and the CXB are shown, respectively. 
Being un-absorbed, the LHB component provides the same contribution in both spectra, while the contribution from the other components is modulated by the best fit column density of absorbing material (Fig. \ref{spec}). 
The best fit column density of absorbing material is significantly different in the two spectra, resulting to be $Log(N_{H_b}/1cm^{-2})=20.47\pm0.03$ and $Log(N_{H_d}/1cm^{-2})=20.83\pm0.02$, respectively. 
In particular, we observe that the model provides an acceptable description of the data with $\chi^2=357.2$ for 325 dof. 
Additionally, we further verify that the best fit column densities derived from the HI4PI map (HI4PI Collaboration 2016) and averaged over the area from which the spectra are accumulated are $Log(N_{H_b}/1cm^{-2})=20.44$ and $Log(N_{H_d}/1cm^{-2})=20.54$, respectively, in relatively good agreement with the values measured from the X-ray spectra. 
Therefore, this corroborates our previous inference that the observed patchiness is induced primarily by the spatial variations of absorbing column densities. 

To further verify this, we leave the normalisation of the CGM component free to vary and obtain consistent best fit values to $\sim25$~\% at 1~$\sigma$, corresponding to $N_{halo_b}=49\pm6$ and $N_{halo_d}=62\pm9$~pc~cm$^{-6}$, respectively. 
No significant improvement in the fit is observed with $\Delta\chi^2=2.9$ for the addition of 2 new degrees of freedom. 
A similar result is obtained by letting the normalisation of the CXB component to be free.  
Indeed, we obtain consistent best fit values to $\sim10$~\% at 1~$\sigma$, corresponding to $N_{CXB_b}=0.028\pm0.009$ and $N_{CXB_d}=0.026\pm0.009$~keV$^{-1}$~cm$^{-2}$~s$^{-1}$~at 1 keV, respectively. 
Finally, a similar result is also obtained by letting the normalisation of the coronal component to be free. 
Indeed, we obtain consistent best fit values to $\sim50$~\% at 1~$\sigma$, corresponding to $N_{Coro_b}=(1.0\pm0.3)$ and $N_{Coro_d}=(0.6\pm0.3)\times10^{-3}$~pc~cm$^{-6}$, respectively. 

This demonstrates that the observed patchiness can be entirely attributed to the variation of the absorber and that the emission measure of the warm X-ray CGM component varied by less than $\sim25$~\% between the bright and dim patches. 

\section{Distance to the absorbing clouds}

\begin{figure}[t]
\begin{center}
\includegraphics[width=0.48\textwidth]{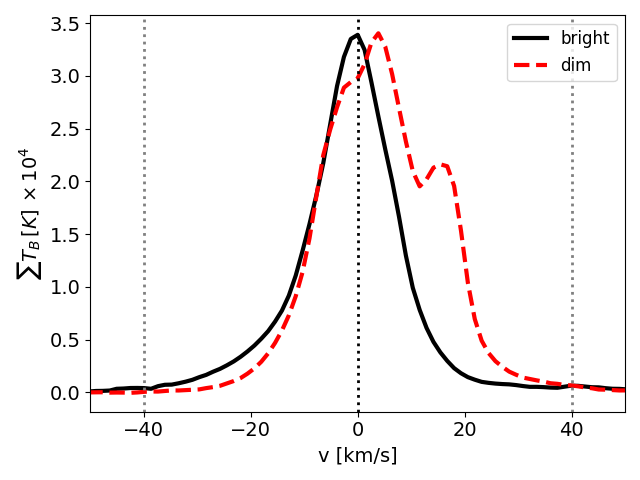}
\caption{Kinematics of the HI absorption towards the bright and dim regions are shown with the black and red lines, respectively. Zero velocities correspond to the velocity of the Sun. }
\label{kine}
\end{center}
\end{figure}

Figure \ref{kine} shows the kinematic of the absorbing clouds towards the bright and dim regions in black and red, respectively, as derived from the HI4PI data cube (HI4PI Collaboration 2016). 
The kinematic of the clouds in the bright and dim regions are different (Fig. \ref{kine}). 
This suggests that, indeed, the two lines of sight are associated with different complexes of absorbing clouds. 
Additionally, the absorption towards the bright region peaks at zero velocities, therefore displaying kinematics consistent with the one of the Sun. 

\begin{figure}[t]
\begin{center}
\includegraphics[width=0.48\textwidth]{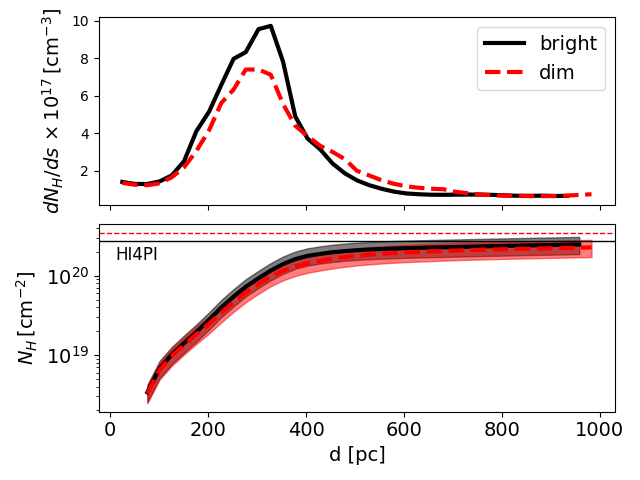}
\caption{The observed extinction as a function of distance towards the bright and dim regions are shown with the black and red curves, respectively. 
Most of the absorption towards the dim region is located beyond $\sim250$~pc from the Sun. }
\label{lal}
\end{center}
\end{figure}

The top panel of Fig. \ref{lal} shows the extinction observed in the spectra of stars, as a function of distance from the Sun (Lallement et al. 2019). 
We take the measurement of the extinction from the Lallement et al. (2019) datacube, which have been computed and averaged from the spectra of a large number of stars with known distance (as measured by \gaia). 
Along both directions, most of the extinction occurs within $\sim200$ and $\sim400$~pc from us. 
By assuming the extinction law from Zhu et al. (2017): $N_H/A_V = (2.08 \pm 0.02) \times 10^{21}$ H cm$^{-2}$ mag$^{-1}$, where $N_H$ is the Hydrogen column density and $A_V$ is the optical extinction, we integrated the extinction and estimated the cumulative column density of absorbing material, as a function of distance from the Sun (bottom panel of Fig. \ref{lal}). 
The error band around the relations in Fig. \ref{lal} reflects the uncertainties on the abundances assumed in the extinction law. 
The bottom panel of Fig. \ref{lal} shows that, within the errors, all the column density of absorbing material is contained within $\sim500$~pc from the Sun, along the bright region. 

On the other hand, the total extinction along the dim region integrated over several hundred parsecs from the Sun is slightly smaller than expected (Fig. \ref{lal}; Lallement et al. 2019), suggesting that part of the absorption might be associated with clouds located further away than several hundred parsecs, therefore most likely towards the upper layers of the HI Galactic disc\footnote{Indeed, the kinematic of such cloud is within $\sim15$~km s$^{-1}$ from the one of the disc. }. 

\section{Discussion} 

\subsection{Comparison with previous results}

By comparing the HaloSat spectra obtained along different lines of sight, Kaaret et al. (2020) observed emission measures of the CGM component which are different by a factor of $\sim2$ on large spatial scales at similar angular distance from the Galactic center (see Fig. 2 of Kaaret et al. 2020). 
In particular, they observed the emission measure of the hot CGM to decrease from values of $EM\sim0.040$~cm$^{-6}$~pc at angular distances from the Galactic center of $\sim40^\circ$ to $EM\sim0.008$~cm$^{-6}$~pc at angular distances of $\sim120-130^\circ$ (see Fig. 2 of Kaaret et al. 2020). 
Additionally, to obtain a good fit (with reduced $\chi^2\sim1$) of the HaloSat data with a simple model (representing emission from either a corona or a halo), Kaaret et al. (2020) added an additional component to the measurement uncertainties with a fixed amplitude of $\sigma_p=0.0034$~cm$^{-6}$~pc. 
Kaaret et al. (2020) showed that such additional scatter is required to fit the HaloSat data, therefore demonstrating that the hot CGM is clumpy. 

Over the entire eFEDS field, located at angular distances from the Galactic center of $d_a\sim120-130^\circ$, we observe a normalisation of the CGM component,  
which corresponds to an emission measure of $EM_{CGM}\sim0.051$~cm$^{-6}$~pc. 
This value is almost one order of magnitude larger than the emission measure derived from the HaloSat data, at similar angular distances from the Galactic center ($EM\sim0.008$~cm$^{-6}$~pc). 
This major difference is the result of the different deconvolution of the observed spectrum of the CGM component. 
Indeed, Kaaret et al. (2020) used a single thermal component to describe the emission of the CGM, which resulted into a best fit temperature of $\sim0.20-0.26$~keV. 
Instead, thanks to the analysis of the higher signal to noise spectrum of the \efeds\ region observed by \erosita, it has been demonstrated that at least two components are needed to fit the CGM emission (Ponti et al. 2022): i) a significantly hotter ($kT\sim0.7$~keV) coronal component, with a much smaller emission measure of $EM\sim0.0009$~cm$^{-6}$~pc; as well as a significantly colder ($kT\sim0.15$~keV) CGM component, with a significantly larger emission measure of $EM\sim0.051$~cm$^{-6}$~pc. 
To verify whether the different emission measures are the result of the different model assumptions, we fitted the bright and dim regions with a single thermal component ({\sc apec}), with its temperature fixed to $kT=0.24$~keV, as observed by Kaaret et al. (2020). 
Despite this model is not statistically acceptable ($\chi^2=495.7$ for 324 dof), it is instructive to notice that the best fit emission measure results to be $EM\sim0.01$~cm$^{-6}$~pc, therefore consistent with the ones derived from the HaloSat data. 
This confirms that the larger emission measure that we observe for the warm X-ray CGM is the result of our fit of the CGM with two thermal components. 

Regardless of the absolute value of the emission measure, the analysis of the HaloSat data indicates an intrinsic variation of the CGM emission by $\frac{\Delta EM}{EM}\sim0.0034/0.010\sim40$~\% at angular distances larger than $\sim120^\circ$ from the Galactic center (Kaaret et al. 2022). 
We note that such result is solid. 
Indeed, the effects of different absorption column densities along different lines of sight are taken into account by Kaaret et al. (2020). 
Therefore, their best fit emission measures of the CGM are de-convolved by the effects of absorption, that we highlight here (Kaaret et al. 2020). 

We also note that an intrinsic patchiness of the CGM of the order of $\sim40$~\% is in slight tension with the upper limits on the emission measure variations of the warm X-ray CGM that we infer along the bright and dim patches within the eFEDS region ($EM_{warm CGM}<25$~\%). 
Additionally, the correlation between the soft X-ray surface brightness and the column density of absorbing material (Fig. \ref{nhcorr}) suggests that the majority of the surface brightness fluctuations (with up to $\sim60$~\% amplitudes) are induced by absorption, again in slight tension with an intrinsic patchiness of the warm X-ray CGM of the order of $\sim40$~\%. 

It is likely that the patchiness of the CGM observed by HaloSat is induced by our hotter CGM component, which we call "corona", for which variations as large as $\sim60$~\% are allowed by the eFEDS data. 
Additionally, it is likely that the patchiness is becoming more prominent when comparing lines of sight beyond the $\sim140$ square degrees of the eFEDS field. 
Other more subtle effects might also play a role. 
For example, spatially non-uniform absorption requires special care to be modeled in spectra of diffuse emission from extended areas (Locatelli et al. 2022). 
The different ions of the Solar wind might have stochastic variations inducing line ratios significantly different from the ones typically employed to model the contribution due to SWCX, among other effects. 

Despite it remains unclear to which CGM component and-or spatial scale the observed patchiness applies, at the end of the all sky survey, \erosita\ is expected to produce a deep exposure of the full sky (Predehl et al. 2021). 
A deep investigation of such dataset will allow us to constrain the level of patchiness at angular scales going beyond the region covered by eFEDS, therefore further clarify these issues. 

\subsection{Modulation of the soft X-ray background over the entire sky}

\begin{figure*}[t]
\vspace{-5.5cm}
\includegraphics[width=1.1\textwidth]{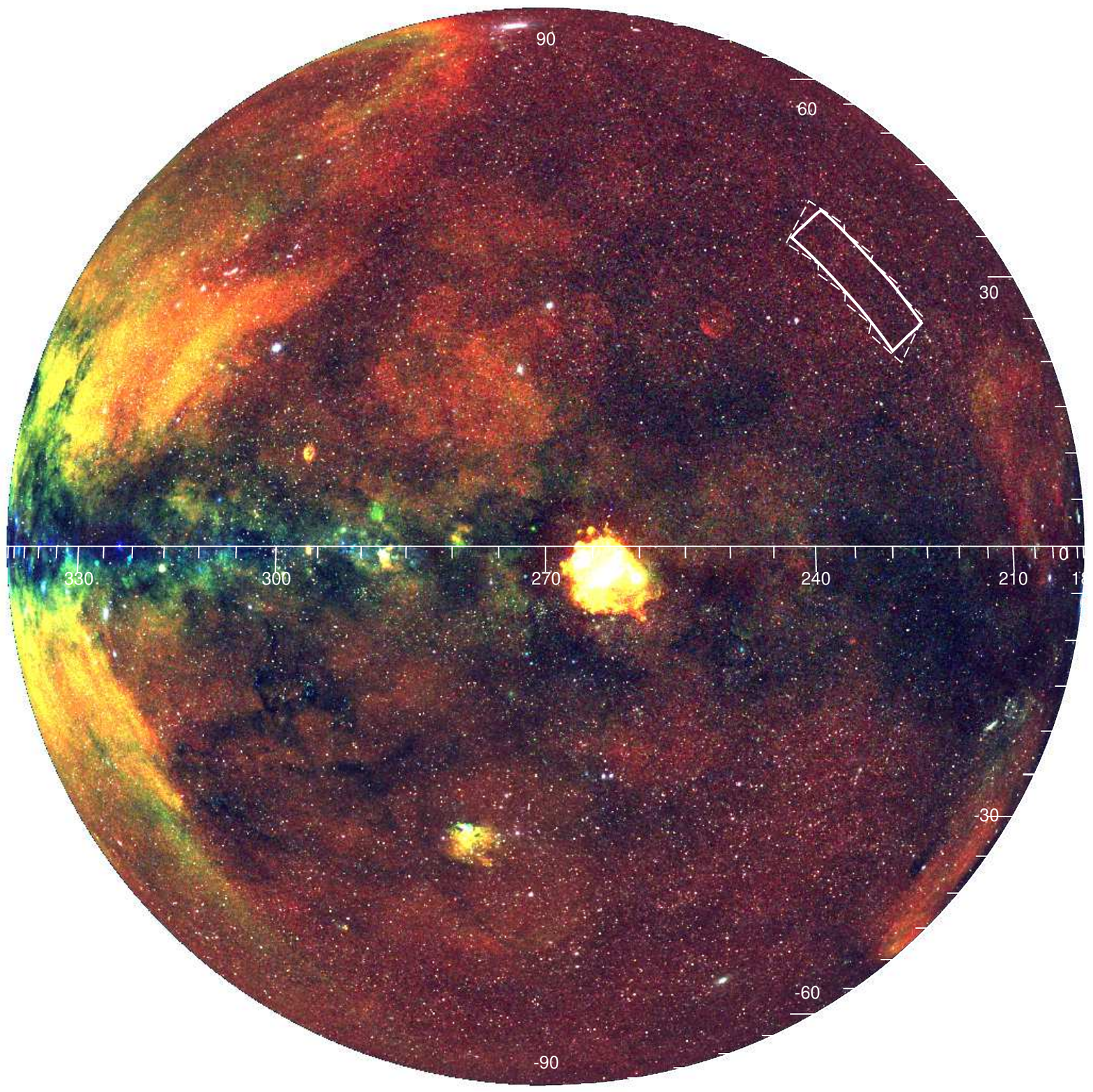}
\vspace{-5.7cm}
\caption{\erosita\ e1 map of the western Galactic hemisphere in orthographic projection. Red, green, and blue show the 0.3--0.6, 0.6--1.0, 1.0--2.3~keV energy bands, respectively. The white regions show the footprint of the \efeds\ region the extraction region of the average \efeds\ spectrum (see Ponti et al. 2022). Patchiness of the soft X-ray diffuse emission in the \efeds\ region can be squinted also in this all sky map. Additionally, this map reveals that similar patchiness away from the Galactic plane and Galactic center. }
\label{allsky}
\end{figure*}

\begin{figure}[t]
\vspace{-3.5cm}
\begin{center}
\includegraphics[width=0.54\textwidth]{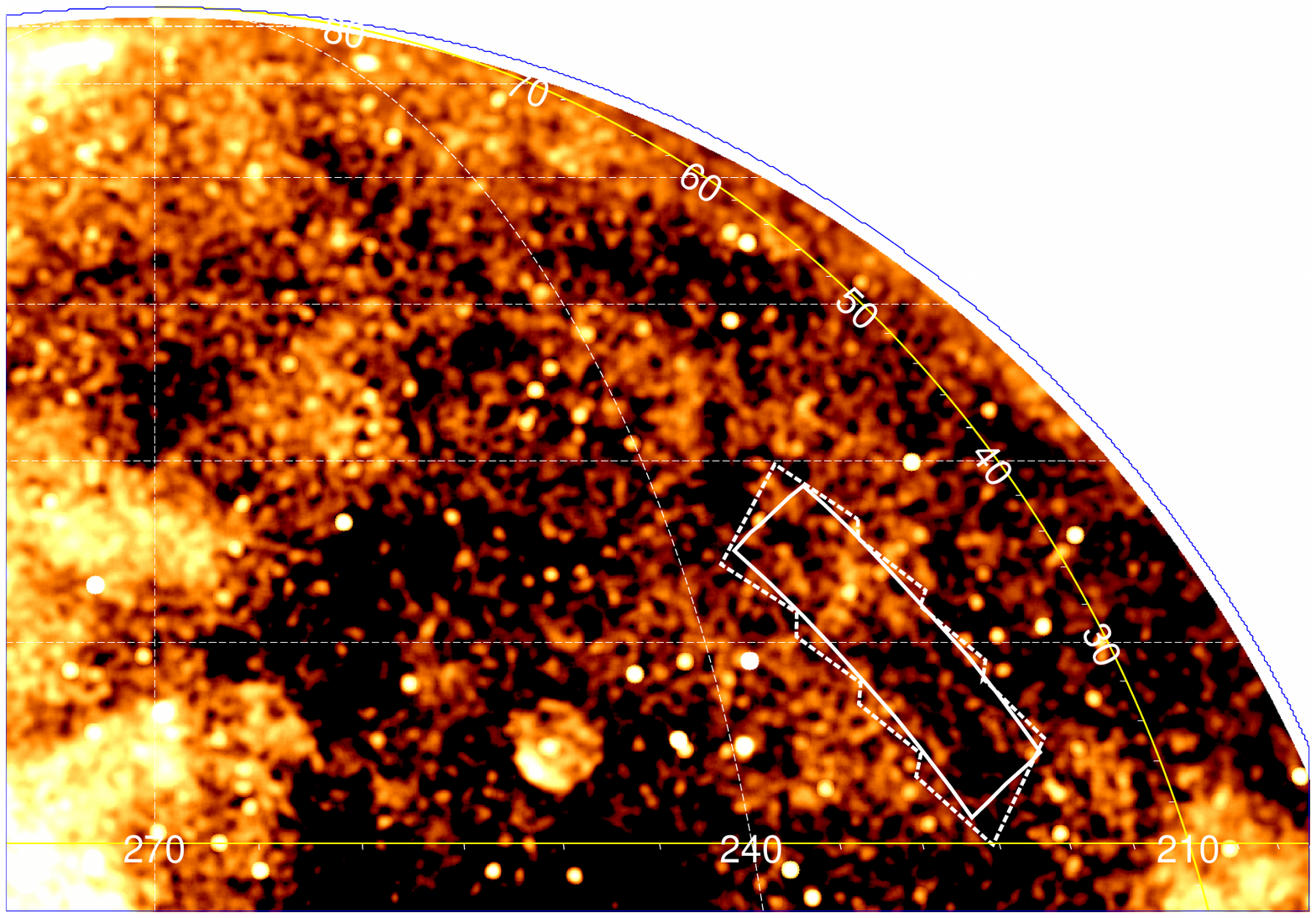}
\vspace{-3.7cm}
\caption{Zoom in of the soft X-ray (0.3--0.6~keV) emission observed towards the north-western Galactic hemisphere. Similar intensity modulations, as observed in the \efeds\ region, are detected over the entire north-western hemisphere. The color scale is shown within a small range to highlight the observed patchiness of the emission. }
\label{CGMPatchy}
\end{center}
\end{figure}

Figure \ref{allsky} shows the \erosita\ map of the western Galactic hemisphere obtained after the completion of the first all sky survey. 
The white lines and the rectangle show the footprint of the \efeds\ region and the extraction region of the \efeds\ spectrum, respectively (Ponti et al. 2022). 
Even in this large scale map, it is possible to observe the patchiness in the brightness of soft X-ray (0.3--0.6~keV) diffuse emission within the \efeds\ region. 

Figure \ref{CGMPatchy} shows a zoom in of the soft X-ray (0.3-0.6~keV) emission towards the north-western hemisphere. The color scale is shown between a small range to highlight the observed patchiness.

A close inspection of these maps (Fig. \ref{allsky} and \ref{CGMPatchy}), reveals that such patchiness is not confined uniquely to the \efeds\ region. 
Instead, it forms a web of brighter and dimmer regions, with amplitudes similar to the one observed in \efeds, over the entire sky\footnote{Clearly this does not apply to the regions towards the Galactic plane and the Galactic center, where the presence of: i) the Galactic outflow; ii) the emission from sources in the plane; iii) the obscuration from large column densities of absorbing material along the plane; etc are the dominant effects.}. 
This suggests that modulations by absorption might be present at other locations in the soft X-ray sky. 

A detailed study of the shadows imprinted on the soft X-ray map by clouds of known distance will be very useful to allow us to place lower limits on the distance from which the bulk of the soft X-ray emission is produced. 
Additionally, it will be important to verify whether any correlation exists between the observed patchiness and the distribution of intermediate velocity clouds as well as the distribution of absorbing Oxygen six features (Tripp et al. 2022). 
Such comparison is critical to understand the origin of the observed patchiness and which fraction is related with intrinsic variations of the diffuse emission. 
Soon, thanks to the detailed analysis of the \erosita\ all sky data, it will be possible to perform such comparison over the entire sky, therefore allowing us to verify the intrinsic CGM patchiness on spatial scales going beyond the $\sim140$ square degrees probed here. 

\section{Conclusions} 

We have investigated the patchiness of the soft X-ray diffuse emission within a contiguous sky region of $\sim140$~square degrees. 

\begin{itemize}
\item{} The soft X-ray diffuse emission in the \efeds\ field shows surface brightness modulations (patchiness) by a factor of $\sim1.6$ on degrees scales.

\item{} The \erosita\ all sky map reveals that similar patchiness in the 0.3--0.6 keV band can be observed, on spatial scales from degrees to tens of degrees, over most of the extragalactic sky, away from the Galactic outflow. 

\item{} We have demonstrated that the observed patchiness can be attributed (within the \efeds\ field) to the known variations in the column density of intervening matter.

\item{} We have shown that Solar wind charge exchange and the hot baryons in Cosmic filaments do not provide a strong contribution to the observed patchiness. However, these processes will have an effect on the patchiness of the soft X-ray diffuse emission along other directions, spatial scales and-or at other times. 

\item{} We have verified that, once the effects of absorption are taken into account, no significant variation of the warm X-ray CGM emission measure is required to fit the data of the bright and dim regions (with an upper limit of $\sim25$~\% on the variation of the emission measure). Therefore, the fit of the soft X-ray emission in the \efeds\ field does not require any intrinsic patchiness of the CGM. 
This can be reconciled with the clumpiness of the CGM observed by Kaaret et al. (2020) either if the latter is induced by the coronal component or if the clumpiness becomes stronger by comparing the CGM properties along lines of sights spreading over the entire sky, instead of being limited to the $\sim140$ square degrees of the \efeds\ regions. 

\item{} Our results highlight the importance of developing proper models to correct for the effects of absorption on X-ray data (i.e., the {\sc disnht} model; Locatelli et al. 2022). 

\item{} These results confirm that our spectral decomposition of the soft X-ray background appears accurate, predicting that, apart from the LHB, all other spectral components are modulated by absorption. 

\end{itemize}

\section{Acknowledgements}

This work is based on data from \erosita, the soft X-ray instrument aboard SRG, a joint Russian-German science mission supported by the Russian Space Agency (Roskosmos), in the interests of the Russian Academy of Sciences represented by its Space Research Institute (IKI), and the Deutsches Zentrum f\"ur Luft- und Raumfahrt (DLR). The SRG spacecraft was built by Lavochkin Association (NPOL) and its subcontractors, and is operated by NPOL with support from the Max Planck Institute for Extraterrestrial Physics (MPE).
The development and construction of the \erosita\ X-ray instrument was led by MPE, with contributions from the Dr. Karl Remeis Observatory Bamberg \& ECAP (FAU Erlangen-Nuernberg), the University of Hamburg Observatory, the Leibniz Institute for Astrophysics Potsdam (AIP), and the Institute for Astronomy and Astrophysics of the University of T\"ubingen, with the support of DLR and the Max Planck Society. The Argelander Institute for Astronomy of the University of Bonn and the Ludwig Maximilians Universit\"at Munich also participated in the science preparation for \erosita.
The \erosita\ data shown here were processed using the eSASS software system developed by the German \erosita\ consortium.

This project acknowledges funding from the European Research Council (ERC) under the European Union’s Horizon 2020 research and innovation programme (grant agreement No 865637). MS acknowledges support from the Deutsche Forschungsgemeinschaft through the grants SA 2131/13-1, SA 2131/14-1, and SA 2131/15-1.

\end{document}